\documentclass{article}

\usepackage{arxiv}

\usepackage[utf8]{inputenc} 
\usepackage[T1]{fontenc}    
\usepackage{hyperref}       
\usepackage{url}            
\usepackage{booktabs}       
\usepackage{amsfonts}       
\usepackage{nicefrac}       
\usepackage{microtype}      
\usepackage{lipsum}		
\usepackage{amsthm,amsmath}
\usepackage{graphicx}
\usepackage[sort, numbers]{natbib}
\usepackage{doi}
\usepackage[utf8]{inputenc} 
\usepackage{textgreek} 
\usepackage{bm} 
\usepackage{graphicx}
\usepackage[separate-uncertainty=true,multi-part-units=single]{siunitx}
\usepackage{cleveref}
\usepackage{csquotes}
\usepackage{booktabs}

\title{Fast and precise model calculation for KATRIN using a neural network}

\author{Christian Karl
    \thanks{also: Physics Department, Technical University Munich,
Garching, Germany} \\
    Max-Planck-Institute for Physics\\
    Munich\\
    Germany\\
	\texttt{karlch@mpp.mpg.de} \\
	\And
	Philipp Eller 
	\thanks{also: Exzellenzcluster ORIGINS, Boltzmannstr. 2, Garching, Germany}\\
	Physics Department\\
    Technical University Munich\\
    Garching\\
    Germany\\
	\texttt{philipp.eller@tum.de} \\
    \And
    Susanne Mertens $^*$\\
    Max-Planck-Institute for Physics\\
    Munich\\
    Germany\\
	\texttt{mertens@mpp.mpg.de} \\
}

\renewcommand{\shorttitle}{KATRIN Neural Net}

\hypersetup{
pdftitle={A template for the arxiv style},
pdfsubject={q-bio.NC, q-bio.QM},
pdfauthor={David S.~Hippocampus, Elias D.~Striatum},
pdfkeywords={First keyword, Second keyword, More},
}

\begin{document}
\maketitle

\begin{abstract} 
    We present a fast and precise method to approximate the physics model of the Karlsruhe Tritium Neutrino (KATRIN) experiment using a neural network.
    KATRIN is designed to measure the effective electron anti-neutrino mass $m_\nu$ using the kinematics of \textbeta-decay with a sensitivity of \SI{200}{\milli \electronvolt} at \SI{90}{\percent} confidence level.
    To achieve this goal, a highly accurate model prediction with relative errors below the \num{e-4}-level is required.
    Using the regular numerical model for the analysis of the final KATRIN dataset is computationally extremely costly or requires approximations to decrease the computation time.
    Our solution to reduce the computational requirements is to train a neural network to learn the predicted \textbeta-spectrum and its dependence on all relevant input parameters.
    This results in a speed-up of the calculation by about three orders of magnitude, while meeting the stringent accuracy requirements of KATRIN.
\end{abstract}


\newcommand{\todo}[1]{{\color{red}{#1}}} 
\newcommand{\comment}[1]{{\color{blue}{#1}}} 

\newcommand{\myparagraph}[1]{\paragraph{#1}\mbox{}\\}

\newcommand{\systparams}{\bm{\theta}_\text{syst}} 
\newcommand{\statparams}{\bm{\theta}_\text{stat}} 
\newcommand{\specparams}{\bm{\theta}_\text{spec}} 
\newcommand{\params}{\bm{\theta}} 
\newcommand{\nll}{-\log\mathcal{L}(\params)} 

\newcommand{\mfit}[3]{$m_\nu^2 = #1_{-#2}^{+#3}\,\si{\electronvolt \squared}$} 

\section{Introduction}

In particle physics, a widely used measurement technique is based on parameter inference using a parameterized model on a set of observed data. 
To obtain precision results, the prediction accuracy of this model, given any choice of parameters, should exceed the statistical uncertainty.
In particular for high-precision experiments, such a requirement can easily translate to maximally allowed model errors at the subpercent level.
Such accurate model predictions often entail high computational demands, that---at least in some cases---can render a full analysis unfeasible and resorting to approximations or a reduced scope becomes necessary.

Some examples of high computational demand are the Monte Carlo methods used by the ATLAS and CMS collaborations at the Large Hadron Collider \cite{atlas2021computing,cmscomputing} to calculate their event predictions and the Ice Cube Collaboration to analyze small signals in high-statistics neutrino oscillation experiments \cite{icecube2019computational}.
The particular example of such a high-precision experiment we want to focus on is the \textbf{Ka}rlsruhe \textbf{Tri}tium \textbf{N}eutrino (KATRIN) experiment, designed to measure the effective electron anti-neutrino mass $m_\nu$ with an unprecedented sensitivity of $\SI{200}{\milli \electronvolt}$\footnote{We use natural units setting $\hbar = c = 1$ for better readability throughout this paper.} at \SI{90}{\percent} confidence level (CL) \cite{designreport}. 
This is achieved via a high-precision measurement of the tritium beta decay spectrum close to the endpoint, where the neutrino mass manifests itself as a small shape distortion.
To reach this design goal, each spectral data point will be measured with a statistical uncertainty of less than one per mill.
This, in turn, puts tight constraints on the required accuracy of the model prediction.
The current KATRIN physics model is based on numerical methods including numerical integrations and root searches, making a single model evaluation computationally expensive, taking approximately one CPU second.
For parameter inference---be that Frequentist or Bayesian---this model then typically needs to be evaluated repeatedly for different parameter values. For KATRIN analyses, often many millions of model evaluations are needed to achieve convergence, driving the computational costs into tens or hundreds of CPU years.
Up until now, it was feasible to apply certain approximations, such as reducing the number of parameters by assuming \SI{100}{\percent} correlations.
However, as the statistical precision increases with the upcoming data taking, these approximations may no longer be applicable.

In this article, we introduce the KATRIN physics model, the computational requirements for future analyses, and how current limitations can be overcome. 
We propose a novel approach to approximate the full KATRIN model by a fast neural network (NN), while retaining the necessary accuracy.
The validity of our approach is discussed by applying it to previous data taking campaigns and Monte Carlo data sets modeling the final KATRIN data set.

\section{The KATRIN physics model}

KATRIN is currently the leading experiment using the kinematics of \textbeta-decay for the measurement of the absolute neutrino mass scale.
Based on the first two data taking campaigns, KATRIN set a limit of $m_\nu < \SI{0.8}{\electronvolt}$ (\SI{90}{\percent} CL) \cite{knm2}.
With the full data set of 1000 days, KATRIN targets a sensitivity close to \SI{0.2}{\electronvolt} at \SI{90}{\percent} CL \cite{designreport}.

\begin{figure}
    \centering
    \includegraphics[width=.95\textwidth]{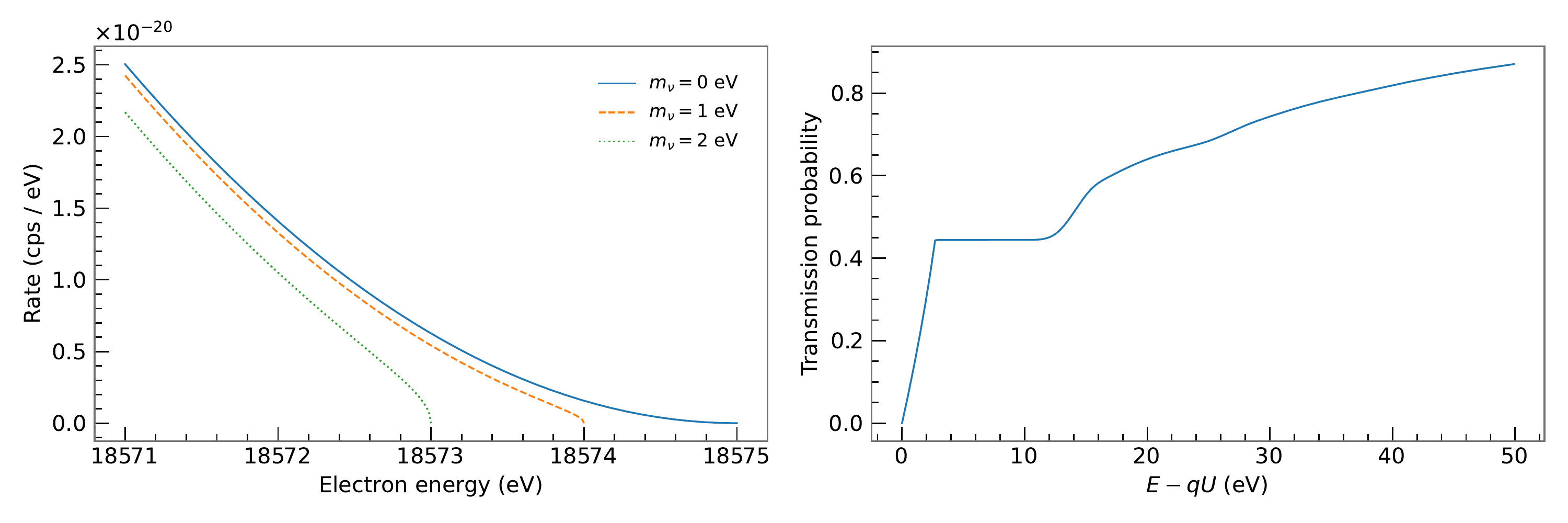}
    \caption{Left: Impact of non-zero neutrino mass on the differential \textbeta-spectrum. The smoking gun is the change in curvature near the endpoint. \\
    Right: Transmission probability of electrons depending on their surplus energy. The sharp edge at low surplus energies is defined by the transmission properties of the MAC-E filter whereas the bumps starting around 13 electronvolt come from energy losses by inelastic scattering of the electron with tritium molecules in the source.
    }
    \label{fig:katrinmodel}
\end{figure}

The measurement principle of KATRIN is based on the \textbeta-decay of tritium \cite{ssc}.
In a \textbeta$^-$-decay a neutron in nucleus $X$ decays into a proton in the daughter nucleus $Y^+$ emitting an electron $e^-$, an electron anti-neutrino $\bar{\nu}$, and the surplus energy $Q$:
\begin{equation}
    X \rightarrow Y^+ + e^- + \bar{\nu} + Q.
    \label{betadecay}
\end{equation}
The surplus energy $Q$ is shared between the decay products with a constant recoil energy $E_\text{rec}$ passed to the daughter nucleus $Y^+$.
The remaining so-called endpoint energy $E_0 = Q - E_\text{rec} = E + E_\nu$ is shared between the electron ($E$) and the neutrino ($E_\nu$).
As $E_\nu = \sqrt{p_\nu^2 + m_\nu^2}$, a non-zero rest-mass of the neutrino reduces the maximal energy the electron can receive.
The differential decay rate for an allowed \textbeta-decay depends on the electron energy and is given by:
\begin{equation}
    \frac{\text{d}\Gamma}{\text{d}E} = C \cdot F(Z', E) \cdot p \cdot (E + m_e) \cdot (E_0 - E) \cdot \sqrt{(E_0 - E)^2 - m_\nu^2} \cdot \Theta(E_0 - E - m_\nu)
    \label{eq:diffspec}
\end{equation}
with a constant term $C$; the Fermi function $F$ with the atomic $Z' = 2$; the electron kinetic energy $E$, momentum $p$ and mass $m_e$; the endpoint energy $E_0$; and the effective neutrino mass $m_\nu$.
The Heaviside function $\Theta$ ensures energy conservation.
The impact of the neutrino mass $m_\nu$ on the spectral shape is maximal near the endpoint $E_0$ as depicted in \cref{fig:katrinmodel} (left).

From \cref{fig:katrinmodel} (left) we can also deduce two of the main properties an experiment aiming to measure $m_\nu$ from \textbeta-decay must fulfill.
First of all, due to the very low rate near the endpoint, an extremely luminous source is required.
Furthermore, an energy resolution on the electronvolt-scale must be achieved to resolve the small shape effect of a non-zero neutrino mass.
To achieve these two features, KATRIN combines a windowless gaseous tritium source \cite{wgts} with an electrostatic high-pass filter with magnetic adiabatic collimation (MAC-E filter) as used by its predecessors in Mainz \cite{mainz} and Troitsk \cite{troitsk}.

As KATRIN uses gaseous molecular tritium as a \textbeta-emitter, the daughter nucleus can end up in a rotational-vibrational or electronic excited state with energy $V_k$ and corresponding probability $P_k$ \cite{saenz}.
This modifies the differential decay rate in \cref{eq:diffspec} as it shifts the maximum energy available for the electron and the neutrino away from the endpoint and smears the spectrum. Introducing $\epsilon_k = E - E_0 - V_k$, we now have:
\begin{equation}
    \frac{\text{d}\Gamma}{\text{d}E} = C \cdot F(Z', E) \cdot p \cdot (E + m_e) \sum_k P_k \cdot \epsilon_k \cdot \sqrt{\epsilon_k^2 - m_\nu^2} \cdot \Theta(\epsilon_k - m_\nu).
    \label{eq:diffspec_fsd}
\end{equation}

Instead of measuring the differential energy spectrum directly, different retarding energies $qU$, where $q$ denotes the electric charge and $U$ the applied voltage, are set in the MAC-E filter and only electrons with sufficient energy to pass the filter are counted resulting in an integral measurement of the \textbeta-spectrum.
The probability of an electron with energy $E$ to pass the retarding energy $qU$ is given by the response function
\begin{equation}
    f(E; qU) = \int \sum_{s=0}^{\infty} T_s(E - \epsilon; qU) \cdot P_s \cdot  g_s(\epsilon) \, \text{d}\epsilon
    \label{eq:response}
\end{equation}
shown in \cref{fig:katrinmodel} (right). Here, the transmission function $T$ describes the properties of the MAC-E filter which are mainly governed by the magnetic fields in the beamline and the angular distribution of electrons in the source, $P_s$ is the probability of an electron to scatter $s$ times, and $g_s(\epsilon)$ describes the probability of an electron to lose the energy $\epsilon$ in the $s$-fold scattering process.

The integral count rate is given by integrating the differential spectrum times the response function over the electron energy:
\begin{equation}
    R(qU) = \int_{qU}^{E_0} \frac{\text{d}\Gamma}{\text{d}E} \cdot f(E; qU) \, \text{d}E.
    \label{eq:intspec}
\end{equation}

Putting all described components together, we can already see why the calculation of the expected rate $R$ is computationally involved.
In the differential spectrum, \cref{eq:diffspec_fsd}, we must sum over hundreds of final states.
To calculate the response function, \cref{eq:response}, we convolve the transmission over the energy loss function for each scattering $s$.
The transmission function itself consists of a numerical root search combined with an integral, and the energy loss function for $s$-fold scattering is obtained by convolving $g$ $s$-times with itself.
Finally, we integrate the differential decay rate over the response function.
Combining all these factors leaves us with a computationally expensive model calculation of about one second for each integral rate.
Although parts can be precalculated depending on the analysis, we will see in the next section that this leads to very high computing demands which are already on the edge of feasibility.

\section{Analysis methods and computing requirements}

The KATRIN collaboration follows multiple approaches to infer the value, uncertainty, and thus confidence interval of the neutrino mass \cite{knm1,knm1prd,knm2}.
In the following, we will describe why each approach requires excessive computational power and would thus benefit from a fast calculation of the model.

\subsection{KATRIN likelihood}

As a basis for our analysis method, we first infer the likelihood function used for KATRIN.
The likelihood $\mathcal{L}(\mu(\params) | \text{data})$ describes how well a set of model parameters $\params$ describes the measured data.
Building on the integrated tritium spectrum from \cref{eq:intspec}, we can write our expected model rate as
\begin{equation}
    \mu(\params; qU) = A_\text{sig} \cdot R(\specparams; qU) + B
    \label{eq:model}
\end{equation}
with a signal normalization $A_\text{sig}$, a constant background rate $B$ and various parameters impacting the tritium spectrum $\specparams$.
A summary of these spectral parameters can be found in \cref{tab:parameters}.

\begin{table}
    \centering
    \caption{Overview of important spectral parameters used in a KATRIN data analysis with their typical value and constraint.
    Parameters without a constraint are always inferred from the data while the constraint is usually used as \SI{1}{\sigma} uncertainty.}
    \label{tab:parameters}
    \begin{tabular}{llcc}
        \toprule
        parameter & description & typical value & typical constraint \\
        \midrule
        $m_\nu^2$ (eV$^2$) & effective electron neutrino mass squared & \num{0.0} & none \\
        $E_0$ (eV) & endpoint of the \textbeta-spectrum & \num{18573.7} & none \\
        $B_\text{src}$ (T) & magnetic field strength in the source & \num{2.52} & \num{0.0063} \\
        $B_\text{max}$ (T) & maximum magnetic field strength & \num{4.24} & \num{0.00423} \\
        $B_\text{min}$ (T) & minimum magnetic field strength & \num{6.3e-4} & \num{6.3e-6} \\
        $\rho d \sigma$ & gas density x scattering cross section & \num{1.364} & \num{0.001} \\
        $\epsilon_i$ &  nine parameters describing the energy loss & $[0.01, 20]$ & $\mathcal{O}(\SI{1}{\percent})$ \\
        $\Delta_\epsilon$ (eV) & longitudinal electric potential asymmetry & \num{0.0} & \num{0.014} \\
        $\sigma_\epsilon^2$ (eV$^2$) & electric potential variance  & \num{0.0182} & \num{0.003} \\
        $\Delta qU$ (eV) & electric potential offset  & \num{-1.907} & fixed \\
        \bottomrule
    \end{tabular}
\end{table}

During a single spectral measurement, called a scan, a fixed set of retarding energies $qU_i$ are scanned with the corresponding measurement times $t_i$, counting the number of electrons $N_i$ that hit the detector.
As the individual measurements at $qU_i$, called scan steps, are statistically independent and the decay rate in the source is constant, we can describe our likelihood using a product of individual Poisson distributions:
\begin{equation}
    \mathcal{L}(\params) = \prod_i P(N_i; \mu(\params; qU_i) \cdot t_i)
    = \prod_i e^{-\mu \cdot t_i} \frac{(\mu \cdot t_i)^N_i}{N_i!}.
\end{equation}

A typical scan contains 40 scan steps, of which about 30 are used for neutrino mass analysis, and takes about two hours.
The KATRIN detector is segmented into 148 pixels, each measuring an independent spectrum.
In simple cases, the pixels can be treated as one detector (uniform), but often segmentation into patches is needed to avoid averaging significantly different models.
Typical segmentations are the grouping into 12 rings or 14 ring-like patches which account for misalignment of the detector with respect to the beamline.
In a measurement campaign, a so-called period in KATRIN jargon, on the order of 300 scans are performed.
For the analysis of a single measurement campaign, one can typically combine the data of all measurements taken at the same $qU$ set points, leading to roughly 30 measurement points for one campaign.
In addition to the detector segmentation, the model can differ significantly between various measurement campaigns.
We must therefore also split our model and data in time, using one model for each campaign.
Some model parameters are individual to each segmentation, while others are shared.
Writing out the total likelihood then leads us to:
\begin{equation}
    \mathcal{L}(\params) = \prod_\text{period} \prod_\text{patch} \prod_i P(N_i; \mu(\params_\text{period,patch}; qU_i) \cdot t_i)
    \label{eq:totallikelihood}
\end{equation}

As an example, we expect roughly 15 periods, each with the 14-patch segmentation, for the total KATRIN measuring time.
This gives us $\approx 15 \cdot 14 \cdot 30 = 6300$ data points and on the order of one thousand likelihood parameters.

\subsection{Parameter inference}

We now describe various methods used to infer these parameters from the likelihood and to obtain an uncertainty on $m_\nu^2$.
In the following, we refer to an approach where only the four unconstrained parameters $\statparams = \{m_\nu^2, E_0, A_\text{sig}, B\}$ are included as \enquote{statistical only}.
We label the constrained \enquote{systematic} parameters $\systparams$.

\myparagraph{Nuisance parameter method}
In this approach, we retrieve the parameters best describing our data by maximizing the likelihood (minimizing the negative log-likelihood) with respect to the parameters $\params$.
To include external constraints, we multiply our likelihood with so-called pull terms.
The simplest and most common form of the pull term is a Gaussian centered at the best knowledge value with a standard deviation describing the uncertainty.
To calculate the uncertainty of $m_\nu^2$, we profile the likelihood searching for the $\Delta \log \mathcal{L}$ value corresponding to $1\,\sigma$.
This is needed, as the errors in $m_\nu^2$ are typically not symmetric and the estimate from the Hesse of the fit is numerically unstable.

Using this approach for all parameters has two important impacts on the computational requirements.
Firstly, the dimensionality of our minimization problem is very large, leading to many likelihood- and thus model evaluations in practice.
In addition, all parts of our model rate, especially the expensive response function, must be recalculated in every minimization step as they depend on $\params$.
Analyzing a single measurement campaign, the detector segmented into 12 or 14 parts, which already requires on the order of \num{e10} evaluations of the integral spectrum when including the required estimation of the derivatives.
With the numerical model, this can take up to one CPU year with drastic parallelization being difficult due to the serial nature of minimization.
The scaling with additional measurement phases is at least quadratic since both the number of points and free parameters increase linearly, making it challenging to use this method as-is for final KATRIN.

\myparagraph{Monte Carlo propagation method}
Another approach used is based upon Monte Carlo propagation of uncertainty.
The general idea is to repeat the maximum-likelihood fit multiple times with randomized input values for the systematic parameters $\systparams$, but to fix them during the minimization.
Statistical uncertainty is included by randomizing the data points according to their underlying Poisson distribution in each step.
The resulting distribution in $m_\nu^2$ describes the uncertainty and can be used to calculate appropriate intervals as well as the best-fit value.

An advantage of this method is that it allows precalculating the expensive response function which depends on $\systparams$, but not on any of the remaining fit parameters $\statparams$.
The downside is that it requires repeating the fit process thousands of times.
This becomes worse when each fit is already expensive, for example when the likelihood has a large number of points due to the necessity of splitting the data over time and over detector segments.
This was already an issue when analyzing the second neutrino mass campaign, where the final result with all uncertainties included more than \num{100000} individual fits, each with \num{e8} function evaluations, amounting to a total of \num{e13}.
With the applied precalculations, this amounted to a computing requirement of about 50 CPU years.
When fitting multiple measurement campaigns together to retrieve the combined neutrino mass value, this quickly becomes unfeasible due to the same quadratic scaling issues as with the nuisance parameter method.

\myparagraph{Full Bayesian sampling}
\label{sub:bayesian}
In addition to the two Frequentist approaches, there is also the option for a full Bayesian analysis.
Instead of performing a minimization, we sample the likelihood using a Markov chain Monte Carlo (MCMC).
External constraints are included by multiplying the likelihood with the corresponding so-called priors.
In this case, similar to the nuisance parameter method, the expensive response function cannot be approximated.
A typical result involves more than \num{e7} samples in the posterior, each with hundreds or thousands of calculations of the integral spectrum and an acceptance of $\approx 0.234$ \cite{acceptancemcmc}, amounting to around \num{e11} total evaluations.
This is more expensive than the nuisance parameter approach and therefore hardly feasible at the current state.

Instead, a model variation approach similar to the Monte Carlo propagation was pursued.
Once again, $\systparams$ are randomized before running a full MCMC chain.
Finally, the samples from all chains are combined to incorporate the effect of systematic uncertainty.
This approximated hybrid method requires at least the computing time of the Monte Carlo propagation method, and can therefore also not be scaled as-is.

\section{Approximating the model estimate with a neural network}

As we have seen in the previous section, all analysis approaches would benefit from a fast calculation of the integrated \textbeta-spectrum $R(qU; \specparams)$, as it is evaluated billions of times.
A possible solution to this is to pre-calculate $R$ for many samples of $\specparams$ and to retrieve $R$ for arbitrary samples within the generated sample range using a multidimensional interpolation algorithm during the analysis.
Unfortunately, the high dimensionality in $\specparams$ of $\mathcal{O}(10)$ as seen in \cref{tab:parameters}, combined with the stringent sub-per mill accuracy requirements, quickly makes traditional interpolation algorithms such as cubic-splines or ($k$-) nearest-neighbor unfeasible.

Our approach to this interpolation problem is to use a neural network: we train the network using samples generated with the analytical model to learn the output rate depending on the physical parameters $\specparams$.
In the following we describe the architecture of the network as well as the exact form of input and output layers we propose to achieve the required accuracy.

\begin{figure}
    \centering
    \includegraphics[width=.95\textwidth]{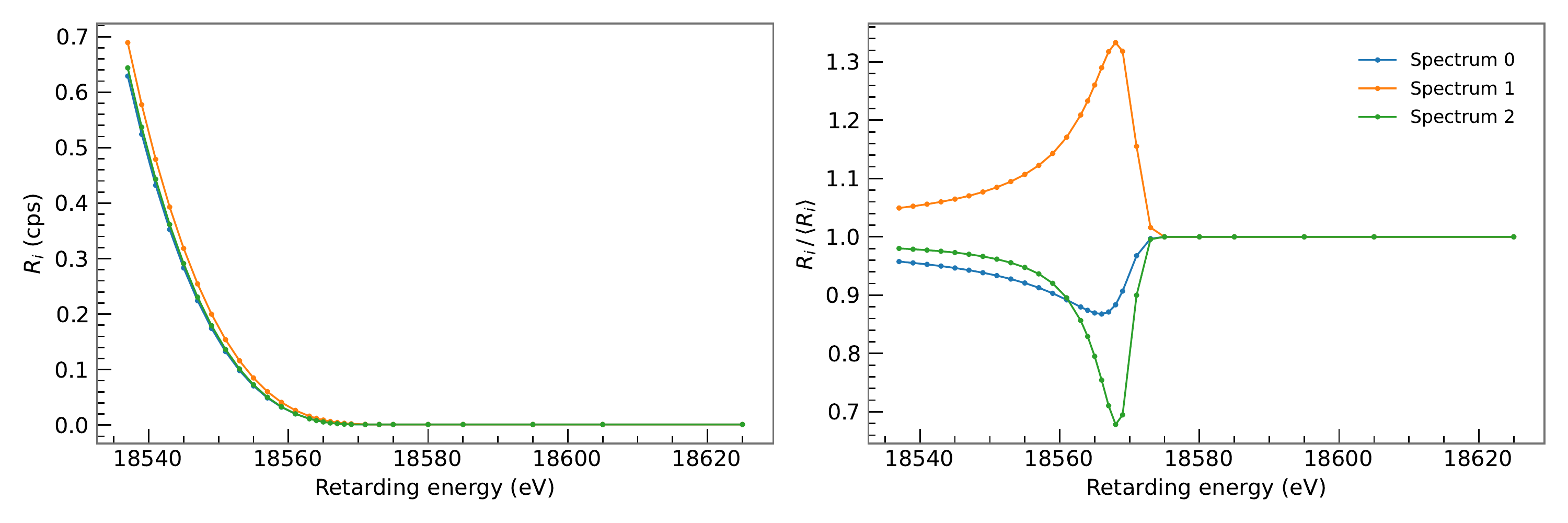}
    \caption{Left: Output spectral rate for different input parameter values $\specparams$. The spectra all show the same general trend. \\
    Right: The output spectra normalized by the average rate in each $qU$ point. We now see clear shape-effects for the different parameter samples.
    }
    \label{fig:netsamples}
\end{figure}

The inputs we pass to the neural network are the values of the parameters $\specparams$.
As we sample these independently, there is no need for decorrelation, but we normalize each parameter to a mean of zero and a standard deviation of one.
As output, we use the predicted count rate $R(qU; \specparams)$ for each point $qU$ in the data spectrum.
This allows the neural network to \enquote{learn} any correlations between the count rate in neighboring points.
As all integrated spectra follow a very similar pattern, see \cref{fig:netsamples} (left), we divide each spectrum by the average count rate at every given point $\left<R(qU; \specparams)\right>$ adding a constant background rate $B$ to avoid dividing by zero.
We then arrive at output samples as shown in \cref{fig:netsamples} (right).

\begin{figure}
    \centering
    \includegraphics[width=.95\textwidth]{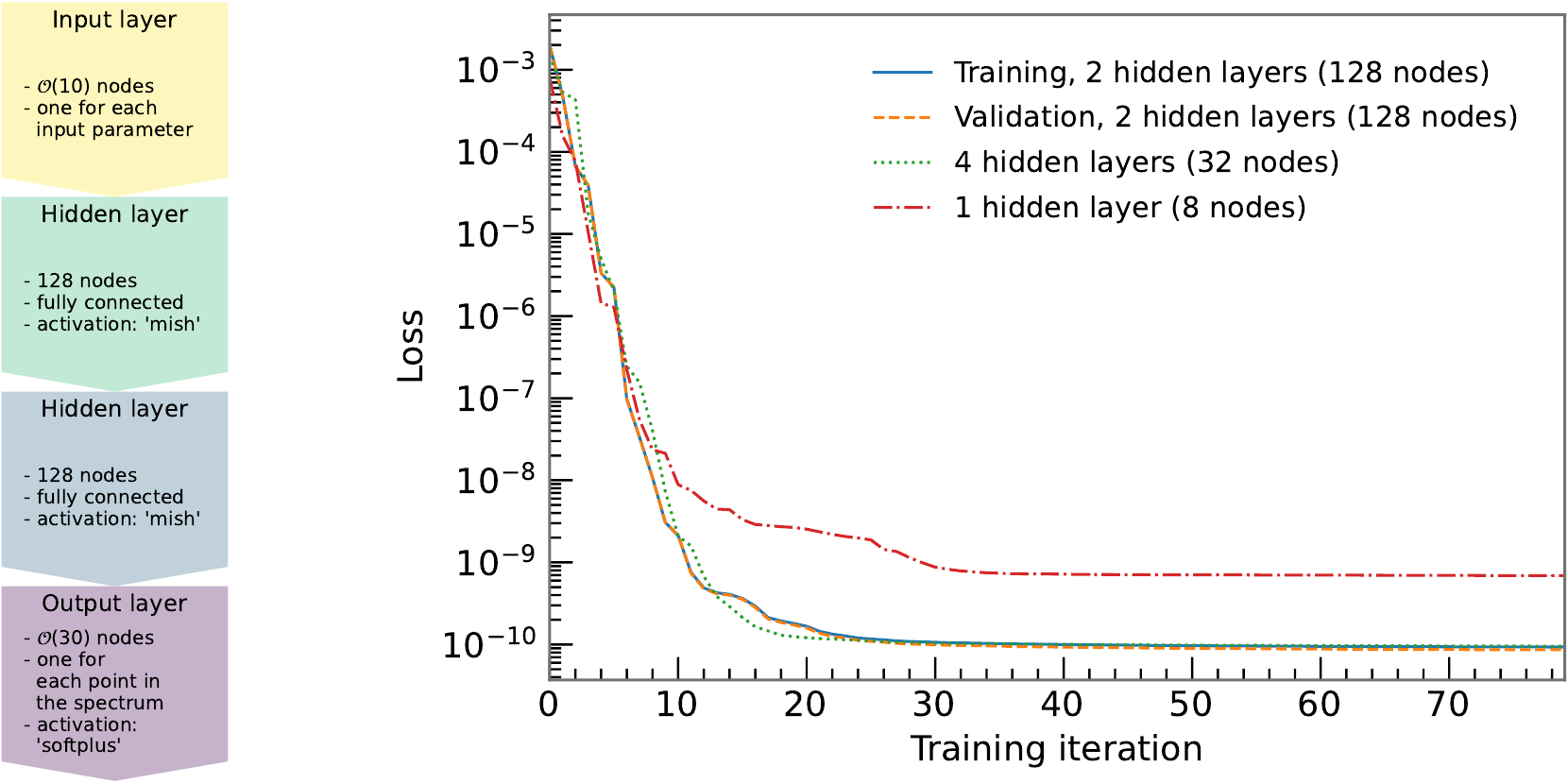}
    \caption{Left: Architecture of the neural network used for model approximation as a block diagram. \\
    Right: Training loss over iteration for the different samples.
    Training (blue solid line) and validation (orange dashed line) are for our reference architecture using two hidden layers with 128 nodes.
    The two match very well, indicating there is no overtraining.
    Including four hidden layers with 32 nodes (green dotted line) has a similar performance to our reference architecture, showing that the exact architecture does not have a large impact on the performance, as long as there are enough weights to optimize.
    Using only one hidden layer with 8 nodes (red dash-dotted line) leads to a significantly worse loss.
    }
    \label{fig:netstructure}
\end{figure}

Between the input and the output layer, we add two hidden layers, each fully connected to both the input/output layer and the next/previous hidden layer.
Each hidden layer uses the \enquote{mish} function as activation and consists of 128 nodes.
As all our rates $R_i$ and thus also the output values $\frac{R_i}{\left<R_i\right>}$ are positive, the output layer uses the \enquote{softplus} activation function.\footnote{
\begin{align*}
    \text{softplus}(x) &= \ln(1 + e^x) \\
    \text{mish}(x) &= x \cdot \tanh(\text{softplus}(x))
    \label{eq:activations}
\end{align*}
}
The final structure of the neural network is depicted in \cref{fig:netstructure} (left).

We train the neural network on $\mathcal{O}(10\,\text{million})$ input samples which we generate by sampling the parameters within their expected 1-, 3- and 5-sigma range using the $N$-dimensional $R_2$ method \cite{r2method}.
During training, we optimize our loss function with respect to the weights of the network using an interface of scipy's \cite{scipy} low-memory BFGS \cite{lbfgs80,lbfgs89} minimizer to the tensorflow \cite{tensorflow} front-end keras.
We define our loss function as the mean squared error of the net prediction
\begin{equation}
    \left<\left(C_\text{i} - C_\text{pred,i}\right)^2\right>
    \label{eq:loss}
\end{equation}
with the true rate change of each sample $C_\text{i} = \frac{R_i}{\left<R_i\right>}$ and the corresponding prediction of the neural net $C_\text{pred,i}$.
Here, we follow an iterative approach.
In each iteration:
\begin{itemize}
    \item we select a random batch of $\mathcal{O}(1\,\text{million})$ samples from our total of $\mathcal{O}(10\,\text{million})$ samples, and
    \item gradually decrease the minimization tolerance.
\end{itemize}
After roughly 40 iterations, we see convergence of our loss function as shown in \cref{fig:netstructure} (right).
During training, we split off \SI{10}{\percent} of our generated samples for validation to check for overfitting.
As the validation loss is the same as the training loss, we conclude that the training is successful and no overfitting happens.

To generate the samples, we need to evaluate the full integral spectrum about $30 \cdot \num{e7} = \num{3e8}$ times, significantly less than for the individual analysis methods.
This task is embarrassingly parallel and the required $< 10$ CPU years can be split upon thousands of CPUs on a computing cluster, allowing us to typically complete this task within a single day.
The training is then no longer challenging using a modern GPU and is completed within a few hours.
One great advantage of this approach is that this computing time must only be invested once and can then be reused for multiple analyses.

\section{Validation and results}

To test the numerical accuracy of our neural network, we do a 1:1 comparison of the analytic model to the NN approximation with statistical uncertainties on representative Monte Carlo data.
Afterwards, we carry out a full analysis of already analyzed measurement phases comparing to the published results.
Finally, we demonstrate that our approach allows the analysis of a large number of datasets as expected for the final KATRIN analysis.

\subsection{Asimov cross-fit}
\label{sub:asimov}

\begin{figure}
    \centering
    \includegraphics[width=0.95\textwidth]{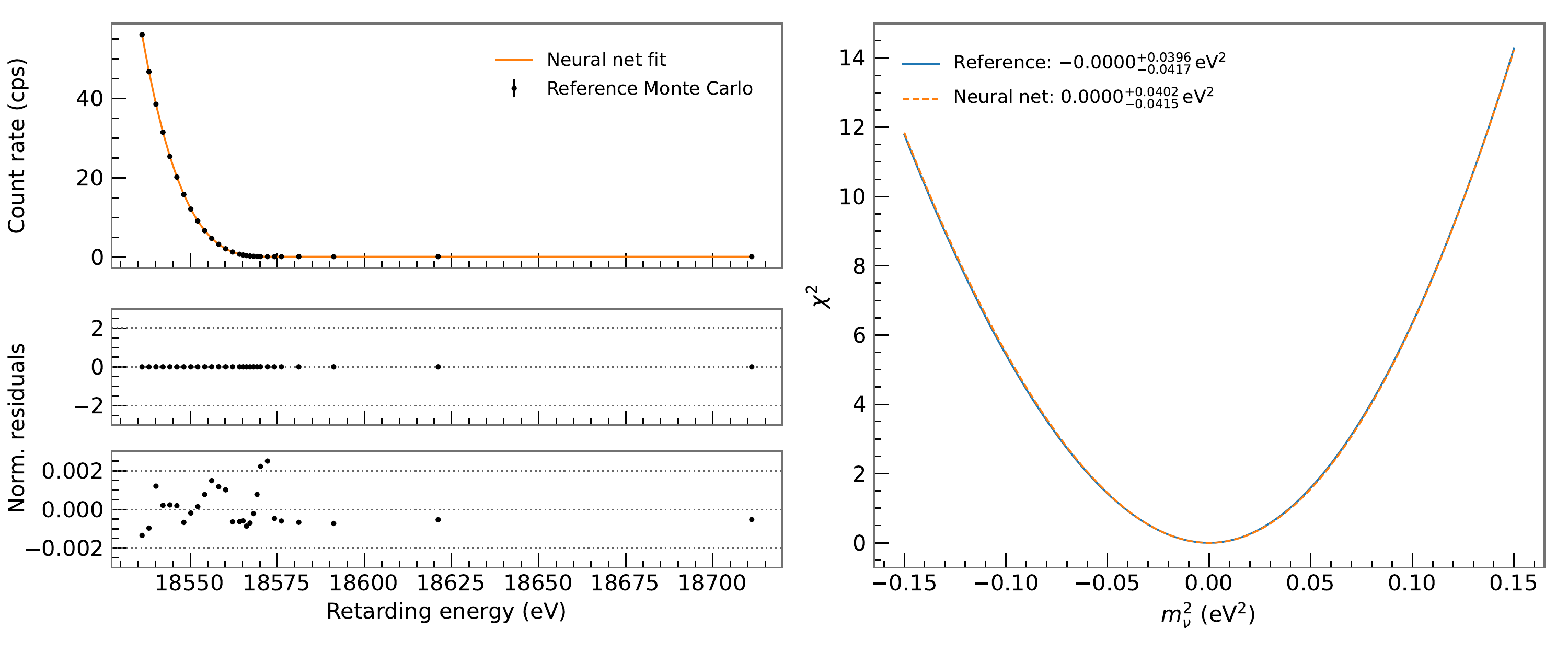}
    \caption{Left: Spectrum and residuals for the Asimov cross-fit.
    On the top we see the generated Monte Carlo spectrum (black points) as well as the best-fit using the neural network (orange line).
    On the bottom we plot the normalized residuals, once with a regular scale, and once with a zoom-in to see the underlying structure. \\
    Right: Comparison of the likelihood-profile in $m_\nu^2$ with statistical uncertainty only.
    The reference (blue solid) matches the profile of the neural network (orange dashed line).
    }
    \label{fig:asimovfit}
\end{figure}

As a first check, we generate a single statistically unfluctuated Asimov reference spectrum using the existing numerical analysis framework assuming 1000 days of measurement.
The parameter values assumed in this spectrum correspond to the ones summarized in \cref{tab:parameters}.
We now fit this spectrum with the model obtained from the NN.
$\statparams$ are treated as free parameters in the fit, while the others are fixed to their input value.
This simplification of fixing $\systparams$ allows a detailed comparison with the existing framework.
The spectrum and residuals are shown in \cref{fig:asimovfit} (left) while the fit parameter results can be found in \cref{tab:asimovfit}.
There is no structure in the residuals when scaled to the usual $1\,\sigma$ level.
Only after zooming in, we can see residuals below the $0.003\,\sigma$ level, almost three orders of magnitude smaller than the expected statistical uncertainty.
This excellent fit is further underpinned by the $m_\nu^2$ bias of less than $\SI{1e-5}{\electronvolt \squared}$.
For the other parameters, we cannot see any deviation on the displayed number of significant digits.
Therefore, we conclude that the neural network has no significant bias when fitting unfluctuated spectra.

In addition to the central value, we also check the likelihood profile in $m_\nu^2$ which is used to retrieve the uncertainty.
The overlay of the two profiles, one using our network, one using the existing methods, is shown in \cref{fig:asimovfit} (right).
We can see that the overlay matches and the $1\,\sigma$ uncertainties differ by less than \SI{1e-3}{\electronvolt \squared}.
Thus, also the uncertainties are not significantly biased using our neural network.

\begin{table}
    \centering
    \caption{Parameter recovery of the Asimov cross-fit. Our neural network recovers all true parameters with the bias in $m_\nu^2$ being less than \SI{1e-5}{\electronvolt \squared}.}
    \label{tab:asimovfit}
    \begin{tabular}{lll}
    \toprule
    fit parameter & true value & recovered value  \\
    \midrule
    $m_\nu^2$ (\si{\electronvolt \squared}) & 0 & \num{-9.8e-6} \\
    $E_0$ (\si{\electronvolt}) & \num{18573.700} &  \num{18573.700} \\
    $A_\text{sig}$ & 1.18 & 1.18 \\
    $B$ (\si{\milli cps}) & 136 & 136 \\
    \bottomrule
    \end{tabular}
\end{table}

\subsection{Ensemble test}

\begin{figure}
    \centering
    \includegraphics[width=0.95\textwidth]{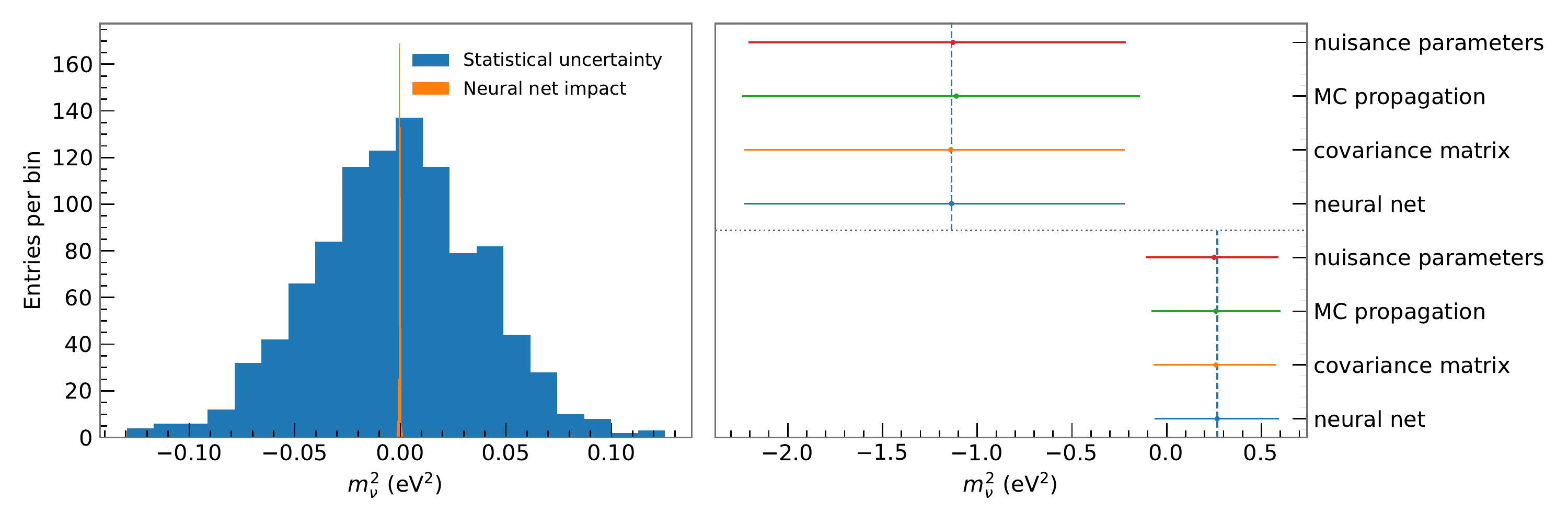}
    \caption{Left: Difference of best-fit $m_\nu^2$ values using the regular analysis framework and our neural network (orange) compared to the assumed statistical uncertainty of \SI{0.04}{\electronvolt \squared} on our 1000 day Monte Carlo spectrum.
    Both bias (\SI{-3e-4}{\electronvolt \squared}) and width (\SI{4e-4}{\electronvolt \squared}) are below the \SI{1e-3}{\electronvolt \squared} level. \\
    Right: Comparison of our analysis of the first two measurement campaigns using the neural network (blue) with the results published in \cite{knm1,knm2}.
    The gray dotted line splits the first measurement campaign (top) and the second one (bottom).
    }
    \label{fig:results}
\end{figure}

As a next step, we perform an ensemble test on 1000 statistically fluctuated spectra using the same 1000 day Asimov spectrum as in \cref{sub:asimov} for the expectation, and the Poisson distribution to randomize it.
We then fit each of these 1000 spectra once with the regular framework and once with the neural network.
In \cref{fig:results} (left) we compare the $m_\nu^2$ best-fit values of the two methods.
The resulting bias (width) in $m_\nu^2$ is \SI{-3e-4}{\electronvolt \squared} (\SI{4e-4}{\electronvolt \squared}) and thus two orders of magnitude smaller than the assumed statistical uncertainty around \SI{0.04}{\electronvolt \squared}.
This would only affect a possible final upper limit at the third significant digit.
We would also like to point out that the numerical inaccuracies of the current analysis methods can also be similarly large.
We thus conclude that the accuracy of the neural network is sufficient to be applied in KATRIN data analysis.

\subsection{Application to realistic KATRIN scenarios}

\myparagraph{Analysis of the first two measurement campaigns}
As a first step, we reproduce the published results of the first two measurement campaigns \cite{knm1,knm2} using the neural network and the nuisance parameter approach.
We include all systematic uncertainties in the official analysis except for an uncertainty on the molecular final states and the source activity fluctuation as these are not parameterized and will be treated differently in future analyses.
Background parameters and uncertainties not shown in \cref{tab:parameters} are included without the use of a neural network as they are simple linear parameters.
For the first measurement campaign, we arrive at \mfit{-1.14}{1.10}{0.91}, consistent with the percent level with the updated value reported in \cite{knm2}.
When analyzing the second measurement campaign using a 12-ring segmentation for the detector, we find \mfit{0.27}{0.33}{0.33} which is within the range of values reported by the analysis teams in \cite{knm2} where the individual results can differ by around \SI{0.01}{\electronvolt \squared}.
The various results both from the published analyses and using our neural network are summarized in \cref{fig:results} (right).

After showing that our method provides consistent results, we can now go beyond what has been published.
To combine the two datasets, we perform a combined fit with shared $m_\nu^2$ using a uniform detector segmentation for the first period, and a 12-ring segmentation for the second measurement phase.
With this approach, we find \mfit{0.07}{0.34}{0.30}.
This type of combined fit has not been performed before, as it was computationally involved and the ring-wise segmentation of the second measurement campaign is not necessarily required.
However, with our neural network-based approach, this fit completes within roughly 10 CPU minutes.
A more detailed overview of the various computing times is shown in \cref{tab:times}.

In addition to the Frequentist approach, we perform a full-Bayesian analysis of the two campaigns sampling both $\statparams$ and $\systparams$ during the MCMC.
As described in \cref{sub:bayesian}, this type of analysis was not feasible with the regular analysis framework.
Using a flat positive prior on $m_\nu^2$, we find consistent limits of $m_\nu < \SI{0.90}{\electronvolt}$ (first campaign), $m_\nu < \SI{0.86}{\electronvolt}$ (second campaign), and $m_\nu < \SI{0.75}{\electronvolt}$ (both combined) at \SI{90}{\percent} credibility.
Each of these analyses was completed in less than one CPU day with a few million samples in the posterior.

We can also see the re-usability aspect of our neural network play out here, as we used the same two trained networks, one per measurement campaign, for each of these analyses.

\myparagraph{Feasibility study for final KATRIN}
Finally, we perform a study to demonstrate the feasibility of analyzing a dataset similar\footnote{Our dataset is similar with respect to the computational demands due to the assumed data segmentation. However, the actual parameter values and uncertainties are still under evaluation by the KATRIN collaboration and our choice is not meant to reflect the final KATRIN sensitivity.} to what is expected for the final 1000 days of KATRIN data.
For this, we generate a Monte Carlo dataset of 15 measurement phases, each with $m_\nu^2 = \SI{0}{\electronvolt \squared}$ as truth and 60 days of measurement time each.
We then perform a combined fit, segmenting the detector into 14 patches and including systematic uncertainties with the nuisance parameter approach and uncertainties similar to \cref{tab:parameters}.
This results in 1336 parameters in the minimization process.
For the neutrino mass squared, we arrive at \mfit{0.000}{0.049}{0.047} after roughly one CPU day of computing time.
The preparation of the required networks, probably one per patch and measurement phase, would take a few weeks at most and can already be done when analyzing the individual campaigns and then reused for all subsequent combined analyses.
This shows that a final combined fit with several hundreds of parameters in the minimization is both numerically and computationally feasible with our approach, thus fulfilling the requirements needed to analyze the expected complete KATRIN data in a simultaneous fit.

\begin{table}
    \centering
    \caption{Computing times for different analyses using the neural network.
    This includes the best-fit as well as an asymmetric error scan for the \SI{1}{\sigma} errors in $m_\nu^2$ including systematic uncertainties with the nuisance parameter approach.
    All computations are run on a single core and should only serve as rough estimates to prove the feasibility of our approach.}
    \label{tab:times}
    \begin{tabular}{ll}
    \toprule
    analysis & computing time (CPU)  \\
    \midrule
    first campaign (uniform) & $< \SI{1}{\minute}$ \\
    second campaign (12-ring) & \SI{4}{\minute} \\
    both combined (uniform - uniform) & \SI{2}{\minute} \\
    both combined (uniform - 12-ring) & \SI{8}{\minute} \\
    both combined (12-ring - 12-ring) & \SI{17}{\minute} \\
    KATRIN final MC (15 x 14-patches) & \SI{1}{\day} \\
    \bottomrule
    \end{tabular}
\end{table}

\section{Conclusion}

We have presented a method to approximate the KATRIN physics model with a neural network.
On a Monte Carlo dataset representing the statistics of 1000 days of KATRIN measurement, we showed that the bias of our model is less than \SI{1e-3}{\electronvolt \squared} and would impact the final limit of KATRIN in the third digit.
Our analysis of the first two data taking campaigns is consistent with the results published on the same level as the individual analysis methods pursued \cite{knm1,knm2}.
Finally, we demonstrated the computational feasibility of performing an analysis of more than 6000 data points and more than 1000 free parameters, emulating the final KATRIN data set.
The presented method will be applied to the forthcoming data taking campaigns of KATRIN, as it offers a solution to overcome the computational limitations of a simultaneous fit of multiple data sets.


\bibliographystyle{unsrtnat}

\section*{Acknowledgements}
We thank the support of the Origins Data Science Lab under the lead of A. Caldwell for making this project possible.
We would also like to thank Prof.~Debarghya Ghoshdastidar from the TUM IT department and his group as well as Ulf Mertens, Sebastian Neubauer, and Felix Wick from Blue Yonder for the fruitful discussions.

We extend our thanks to the full KATRIN collaboration for helpful discussions \& inputs, and finally their review and approval of our work and this article.

\section*{Funding}
We acknowledge the support of Helmholtz Association (HGF), the Max Planck Research Group (MaxPlanck@TUM) program, the international Max Planck School (IMPRS), the Deutsche Forschungsgemeinschaft (DFG, German Research Foundation) for the SFB-1258 program and especially the ORIGINS Data Science Lab (ODSL) of the ORIGINS Excellence Cluster under Germany's Excellence Strategy - EXC-2094 - 390783311, and finally the Munich Data Science Institute (MDSI).
This project has received funding from the European Research Council (ERC) under the European Union Horizon 2020 research and innovation programme (grant agreement No. 852845).
We thank the computing cluster support at the Max Planck Computing and Data Facility (MPCDF).

\section*{Availability of data and materials}
The KATRIN data for the first two measurement campaigns is or will be available via the corresponding publications.
Any other data is available from the corresponding author on reasonable request.

\section*{Competing interests}
The authors declare that they have no competing interests.

\section*{Authors' contributions}

C.~Karl: main author, development of code and application to KATRIN,
P.~Eller: support for neural network development and optimization,
S.~Mertens: supervisor of C.~Karl, initiator of the project.
All authors read and approved the final text.


\bibliography{bmc_article}      

\begin{thebibliography}{18}
\providecommand{\natexlab}[1]{#1}
\providecommand{\url}[1]{\texttt{#1}}
\expandafter\ifx\csname urlstyle\endcsname\relax
  \providecommand{\doi}[1]{doi: #1}\else
  \providecommand{\doi}{doi: \begingroup \urlstyle{rm}\Url}\fi

\bibitem[{ATLAS Collaboration}(2021)]{atlas2021computing}
{ATLAS Collaboration}.
\newblock Atlfast3: the next generation of fast simulation in atlas, 2021.

\bibitem[{CMS Collaboration} and Sekmen(2017)]{cmscomputing}
{CMS Collaboration} and Sezen Sekmen.
\newblock Recent developments in cms fast simulation, 2017.

\bibitem[{IceCube Collaboration} et~al.(2019){IceCube Collaboration}, Aartsen,
  and et~al.]{icecube2019computational}
{IceCube Collaboration}, M.~G. Aartsen, and et~al.
\newblock Computational techniques for the analysis of small signals in
  high-statistics neutrino oscillation experiments, 2019.

\bibitem[{KATRIN Collaboration}(2005)]{designreport}
{KATRIN Collaboration}.
\newblock Katrin design report 2004.
\newblock Technical report, {Forschungszentrum, Karlsruhe}, 2005.
\newblock 51.54.01; LK 01; Auch: NPI ASCR Rez EXP-01/2005; MS-KP-0501.

\bibitem[{KATRIN Collaboration} et~al.(2021{\natexlab{a}}){KATRIN
  Collaboration}, Aker, and et~al.]{knm2}
{KATRIN Collaboration}, M.~Aker, and et~al.
\newblock First direct neutrino-mass measurement with sub-ev sensitivity,
  2021{\natexlab{a}}.

\bibitem[Kleesiek and et~al.(2018)]{ssc}
Marco Kleesiek and et~al.
\newblock $\beta$-decay spectrum, response function and statistical model for
  neutrino mass measurements with the katrin experiment, 2018.

\bibitem[Heizmann et~al.(2017)Heizmann, Seitz-Moskaliuk, and {KATRIN
  collaboration}]{wgts}
Florian Heizmann, Hendrik Seitz-Moskaliuk, and {KATRIN collaboration}.
\newblock The windowless gaseous tritium source (wgts) of the katrin
  experiment.
\newblock \emph{Journal of Physics: Conference Series}, 888\penalty0
  (1):\penalty0 012071, 2017.
\newblock URL \url{http://stacks.iop.org/1742-6596/888/i=1/a=012071}.

\bibitem[Kraus and et~al.(2005)]{mainz}
Ch~Kraus and et~al.
\newblock Final results from phase ii of the mainz neutrino mass search in
  tritium $\beta$ decay.
\newblock \emph{The European Physical Journal C - Particles and Fields},
  40\penalty0 (4):\penalty0 447--468, 2005.
\newblock ISSN 1434-6052.
\newblock \doi{10.1140/epjc/s2005-02139-7}.
\newblock URL \url{https://doi.org/10.1140/epjc/s2005-02139-7}.

\bibitem[Aseev and et~al.(2011)]{troitsk}
V.~N. Aseev and et~al.
\newblock Upper limit on the electron antineutrino mass from the troitsk
  experiment.
\newblock \emph{Phys. Rev. D}, 84:\penalty0 112003, 2011.
\newblock \doi{10.1103/PhysRevD.84.112003}.
\newblock URL \url{https://link.aps.org/doi/10.1103/PhysRevD.84.112003}.

\bibitem[Saenz et~al.(2000)Saenz, Jonsell, and Froelich]{saenz}
Alejandro Saenz, Svante Jonsell, and Piotr Froelich.
\newblock Improved molecular final-state distribution of ${\mathrm{het}}^{+}$
  for the $\mathit{\ensuremath{\beta}}$-decay process of ${T}_{2}$.
\newblock \emph{Phys. Rev. Lett.}, 84:\penalty0 242--245, 2000.
\newblock \doi{10.1103/PhysRevLett.84.242}.
\newblock URL \url{https://link.aps.org/doi/10.1103/PhysRevLett.84.242}.

\bibitem[{KATRIN Collaboration} et~al.(2019){KATRIN Collaboration}, Aker, and
  et~al.]{knm1}
{KATRIN Collaboration}, M.~Aker, and et~al.
\newblock Improved upper limit on the neutrino mass from a direct kinematic
  method by katrin.
\newblock \emph{Phys. Rev. Lett.}, 123:\penalty0 221802, Nov 2019.
\newblock \doi{10.1103/PhysRevLett.123.221802}.
\newblock URL \url{https://link.aps.org/doi/10.1103/PhysRevLett.123.221802}.

\bibitem[{KATRIN Collaboration} et~al.(2021{\natexlab{b}}){KATRIN
  Collaboration}, Aker, and et~al.]{knm1prd}
{KATRIN Collaboration}, M.~Aker, and et~al.
\newblock Analysis methods for the first katrin neutrino-mass measurement.
\newblock \emph{Phys. Rev. D}, 104:\penalty0 012005, Jul 2021{\natexlab{b}}.
\newblock \doi{10.1103/PhysRevD.104.012005}.
\newblock URL \url{https://link.aps.org/doi/10.1103/PhysRevD.104.012005}.

\bibitem[Bédard(2008)]{acceptancemcmc}
Mylène Bédard.
\newblock Optimal acceptance rates for metropolis algorithms: Moving beyond
  0.234.
\newblock \emph{Stochastic Processes and their Applications}, 118\penalty0
  (12):\penalty0 2198--2222, 2008.
\newblock ISSN 0304-4149.
\newblock \doi{https://doi.org/10.1016/j.spa.2007.12.005}.
\newblock URL
  \url{https://www.sciencedirect.com/science/article/pii/S0304414907002177}.

\bibitem[Roberts(2018)]{r2method}
Martin Roberts.
\newblock The unreasonable effectiveness of quasirandom sequences, 2018.
\newblock URL
  \url{http://extremelearning.com.au/unreasonable-effectiveness-of-quasirandom-sequences/}.

\bibitem[Virtanen et~al.(2020)Virtanen, et~al., and {SciPy 1.0
  Contributors}]{scipy}
Pauli Virtanen, et~al., and {SciPy 1.0 Contributors}.
\newblock {{SciPy} 1.0: Fundamental Algorithms for Scientific Computing in
  Python}.
\newblock \emph{Nature Methods}, 17:\penalty0 261--272, 2020.
\newblock \doi{10.1038/s41592-019-0686-2}.

\bibitem[Nocedal(1980)]{lbfgs80}
Jorge Nocedal.
\newblock Updating quasi-newton matrices with limited storage.
\newblock \emph{Mathematics of Computation}, 35\penalty0 (151):\penalty0
  773--782, 1980.
\newblock ISSN 00255718, 10886842.
\newblock URL \url{http://www.jstor.org/stable/2006193}.

\bibitem[Liu and Nocedal(1989)]{lbfgs89}
{Dong C.} Liu and Jorge Nocedal.
\newblock On the limited memory bfgs method for large scale optimization.
\newblock \emph{Mathematical Programming}, 45\penalty0 (3):\penalty0 503--528,
  December 1989.
\newblock ISSN 0025-5610.

\bibitem[Abadi and et~al.(2015)]{tensorflow}
Mart\'{\i}n Abadi and et~al.
\newblock {TensorFlow}: Large-scale machine learning on heterogeneous systems,
  2015.
\newblock URL \url{https://www.tensorflow.org/}.
\newblock Software available from tensorflow.org.

\end{thebibliography}




\end{document}